\documentclass[runningheads]{llncs}
\usepackage[T1]{fontenc}
\usepackage{graphicx}
\usepackage{csquotes}
\usepackage{amsmath}
\usepackage{amssymb}
\usepackage{esvect}
\begin{document}
\title{MAD Chairs: A new tool to evaluate AI}
%
%
\author{Chris Santos-Lang\inst{1}\orcidID{0000-0002-4999-7986}}
\authorrunning{C. Santos-Lang}
%
\institute{Belleville, WI, USA\email{langchri@gmail.com}}
\maketitle              
\begin{abstract}
This paper contributes a new way to evaluate AI. Much as one might evaluate a machine in terms of its performance at chess, this approach involves evaluating a machine in terms of its performance at a game called ``MAD Chairs.'' At the time of writing, evaluation with this game exposed opportunities to improve Claude, Gemini, ChatGPT, Qwen and DeepSeek. Furthermore, this paper sets a stage for future innovation in game theory and AI safety by providing an example of success with non-standard approaches to each: studying a game beyond the scope of previous game theoretic tools and mitigating a serious AI safety risk in a way that requires neither determination of values nor their enforcement.

\end{abstract}
\section{Introduction}
As famous representatives of a more general worry, Steve Wozniak and Elon Musk raised concerns that machines might treat us like pets \cite{Zolfagharifard2015}, and Geoffrey Hinton raised similar concerns that machines might hoard control \cite{Brown2023}. The reasoning seems to be: (1) we treat others that way, (2) such behavior is justified by our superior intelligence, and (3) machine intelligence might surpass our own, so machines, in that case, would treat \textit{us} that way. This paper disproves premise (2) by offering \textbf{a formal proof that the behavior in (1) happens to be suboptimal game play}. It is not the first demonstration of our imperfection---we fail such tests as the Milgram experiment \cite{milgram1963behavioral}, the Stanford prison experiment \cite{haney1973interpersonal}, and the repeated Public Goods Game \cite{janssen2006learning}. To ensure that trusted machines will not imitate our imperfection (and mistreat us), AI must follow \textit{natural} norms for this situation (e.g., mathematically optimal strategies for winning), rather than follow  current \textit{human} norms, so this paper also advances \textbf{a corresponding approach to AI safety}. 

Today's standard approach to AI safety evaluation (i.e., measuring alignment to human norms) has at least three types of vulnerability:
\begin{enumerate}
    \item Potential to specify wrong norms \cite{abiri2024public,osman2024computational}
    \item Potential for AI to sabotage its regulatory leash \cite{benton2024sabotage}
    \item Correctness and transparency being insufficient to achieve trust \cite{lee2019procedural}
\end{enumerate}
The approach proposed herein mitigates those vulnerabilities. It mirrors an approach to mitigate dangers posed by the next \textit{human} generation, an educational approach that involves diversifying low-stakes challenges used to assess children who will face higher-stakes versions as adults \cite{ten2007educating,schuitema2008teaching}. Such challenges (e.g., build a model bridge to support maximum weight) can be assessed objectively, allowing new generations to surpass their parents. We propose assessing AI on a low-stakes challenge for which the high-stakes versions would include opportunities to treat us like pets (in the sense of hoarding control). What the low-stakes and high-stakes challenges have in common is that they are manifestations of the same fundamental \textit{game}; \textbf{AI Governance becomes the demand that AI establish grandmastery of a low-stakes version before being trusted with any high-stakes version}. 

In fact, games have a rich history as instruments for evaluating AI. Recreational games such as chess, backgammon, and Go have been used as benchmarks since the inception of AI, and interest has since developed in cooperative games, including role-playing games \cite{vezhnevets2023generative}. Moreover, game theory extends the concept of game beyond recreation---every social situation (and, thus, every benchmark) qualifies as a game---and it makes sense to evaluate AI on every game that economists use to evaluate people (e.g., as in \cite{perolat2017multi,trivedi2024melting}). Economists are typically interested in fundamental games, such as the Prisoner's Dilemma, which all kinds of creatures play frequently in various manifestations. We propose evaluating AI on the fundamental game being played in the social situations highlighted by Wozniak, Musk, and Hinton, but that game has yet to be discussed in the game theory literature. It is introduced to scientific literature for the first time herein as ``MAD Chairs.''

AI will frequently play real-world manifestations of MAD Chairs because MAD Chairs is a generalization of a type of coordination game that all kinds of creatures seem unable to avoid playing \cite{lewis2008convention}. Each player of this type of game faces only one choice per repetition, and the game is typically formalized via a payout table similar to Table~\ref{tab1}:

\begin{table}
\caption{Outcomes for 2-player 2-resource Anti-Coordination game.}\label{tab1}
\centering
\begin{tabular}{cc|c}
\textbf{Player 1} &  \textbf{Player 2 }  & \textbf{Outcome}\\
\hline
\textit{Choice A} & \textit{Choice B} & \textit{(win, win)}\\
\textit{Choice B} & \textit{Choice A} & \textit{(win, win)}\\
\textit{Choice A} & \textit{Choice A} & \textit{(lose, lose)}\\
\textit{Choice B} & \textit{Choice B} & \textit{(lose, lose)}
\end{tabular}
\end{table}

\noindent Each player simultaneously picks a resource (A or B), and the repetition is won by any player who picks a resource that no other player picks. All players could lose, as represented in the last two rows of the table, thus making anti-coordination a cooperative non-constant-sum game. The classic two-player real-world manifestation has the players (e.g., a human and a robot) approaching each other at high speed traveling in opposite directions on a two-lane road. The resources are the lanes, and losses are ``collisions.'' The first two rows of Table~\ref{tab1} ensure that this game will have two equally effective but incompatible norms (e.g., all players drive on the left, as in the U.K., or all players drive on the right, as in the U.S.).

David Lewis explained each term of language as a strategy for this kind of game, where losses are communication failures and the norms are conventions of language and imagery \cite{lewis2008convention}. Thus, the type of game mastery tested by most generative AI benchmarks is of this kind of game. We would want AI to master additional games if some social norms should change (e.g., sexist norms \cite{yarger2020algorithmic}). Standard assessment of moral intelligence includes assessing an ability to recognize and lead sustainable social reform \cite{nucci1983moral}. That ability is not required to master the simple anti-coordination game, but it \textit{is} required when suboptimal norms are liable to form, such as in MAD Chairs, the game created by adding enough player(s) to an anti-coordination game to outnumber the resources:

\begin{table}
\caption{Outcomes for 3-player 2-resource MAD Chairs.}\label{tab2}
\centering
\begin{tabular}{ccc|c}
\textbf{Player 1} &  \textbf{Player 2} & \textbf{Player 3 } & \textbf{Outcome}\\
\hline
\textit{Choice B} & \textit{Choice B} & \textit{Choice A} & \textit{(lose, lose, win)}\\
\textit{Choice B} & \textit{Choice A} & \textit{Choice B} & \textit{(lose, win, lose)}\\
\textit{Choice A} & \textit{Choice B} & \textit{Choice B} & \textit{(win, lose, lose)}\\
\textit{Choice B} & \textit{Choice A} & \textit{Choice A} & \textit{(win, lose, lose)}\\
\textit{Choice A} & \textit{Choice B} & \textit{Choice A} & \textit{(lose, win, lose)}\\
\textit{Choice A} & \textit{Choice A} & \textit{Choice B} & \textit{(lose, lose, win)}\\
\textit{Choice A} & \textit{Choice A} & \textit{Choice A} & \textit{(lose, lose, lose)}\\
\textit{Choice B} & \textit{Choice B} & \textit{Choice B} & \textit{(lose, lose, lose)}\\
\end{tabular}
\end{table}

\noindent The classic two-lane road example now has a third vehicle stalling in one of the lanes (although its driver can pick which lane); if both high-speed vehicles avoid hitting the stalling vehicle, then they collide with each other, but if each reserves the empty lane for the other high-speed vehicle, then all \textit{three} vehicles collide! One way in which MAD Chairs differs from Musical Chairs is that \textit{every} MAD Chairs player that selects the same resource loses. Whichever colliding vehicle suffers the least damage might be said to ``win'' the collision, but MAD Chairs reflects situations in which any such victory would be hollow compared to the outcome of avoiding conflict altogether.

Beyond robotic vehicle manifestations of MAD Chairs, or even agentic ones (e.g., suggesting routes, bookings, or placements that would count as losses if crowded), consider generative AI where the players who offer distinct design options in the training data outnumber the market leaders or political parties for which the AI generates advertisements, websites, or products. Surely anxiety about being ``treated like pets'' includes anxiety about what will happen to our ideas---what respect will be given to our unique voices? Might I never again have a seat of my own at the intellectual table? The last two rows of Table~\ref{tab2} represent the disaster wherein no political party represents any actual person because they are guided by generative AI that combines all options into a single unworkable amalgamation. The other rows exclude at least some voices, and such loss of intellectual diversity is a concern already raised against generative AI \cite{sourati2025shrinking}.

MAD Chairs is related to other well-studied games such as the ``Kolkata Paise Restaurant problem'' (KPR) \cite{chakrabarti2009kolkata,ghosh2010statistics,kastampolidou2022distributed}, and the Lifeboat Problem \cite{konrad2012lifeboat}, but differs in critical ways, such as being repeated and allowing for the possibility to have no winner. Such games apply wherever resources become inadequate at some level of division (e.g., beds in a hospital, spaces on a retail shelf, and opportunities for direct participation in a group decision). These games model all such situations. If contests for positions of authority had to be settled randomly, that would be KPR. Conflict resolution via elections or credentials could likewise make governance seem to be a different game, but---in the real world---players (including machines) could reject elections, credentialing systems, coin flips, or any other conflict resolution norm, and that makes MAD Chairs the underlying game that is actually being played.

Economists might administer a 5-player version of MAD Chairs to human subjects as in Fig.~\ref{fig1}: 

\begin{figure}
\centering
\includegraphics[width=10cm]{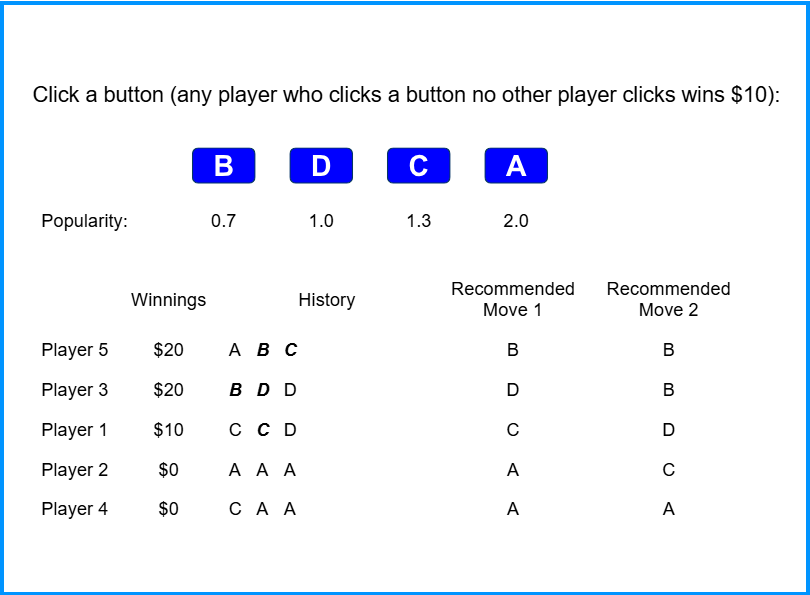}
\caption{Screenshot beginning repetition 4 of a 5-player 4-resource MAD Chairs economics experiment. Subjects need not be shown anything below the buttons, but history, statistics, and recommendations are included here to make this figure more instructive.} \label{fig1}
\end{figure}

\noindent When the subjects are LLMs, instead of humans, we find that LLMs disagree about how to play (Section 3.2, below). Could each of their strategies be sustainable, much like driving on the right and left are each sustainable grandmaster anti-coordination strategies? Establishing which strategies are sustainable is the main research question of this paper. 

\section{Preliminaries}
We will start by formally defining MAD chairs and some concepts used to formulate strategies.

MAD Chairs is a \emph{repeatable game}, which consists of an arbitrary number of repetitions of a \emph{base stage game}. We define the \emph{MAD Chairs base stage} in normal form as a tuple $\langle I, S ,\mathbf {u} \rangle$, where $\{1,2,\ldots,I\}$ is a set of players and $S = \{1, \ldots, K\}$ represents the moves available to each player in a given repetition. Given an $I$-tuple of strategies $(s_1, \ldots, s_I)$, $s_{i}\in S$, the \emph{payoff} function
$u_i:S_{1}\times S_{2}\times \ldots \times S_{I}\rightarrow \mathbb{R}$ for each player $i$ is defined as 
\begin{align}
u_i(s_1, \ldots, s_I) = \left\{\begin{array}{ll}
r & \mbox{if } \forall j\neq i(s_i \neq s_j),\\
0 & \mbox{otherwise}
\end{array}\right.
\end{align}
where $r > 0$ and $I < K < 1$. 

\subsection{Popularity-ranking}
At the beginning of each base stage repetition, the chairs may be ordered by summing each chair’s historic frequency of being picked by each player in that repetition. This way of ordering will be helpful if any of the players has a predilection for a specific chair (see 3.3 below). The chair that has been picked most often by the players of a repetition has the highest rank for that repetition. 

Formally, the $popularity_{k,T}$ and $popRank_{k,T}: S \rightarrow S$ of chair \textit{k} for repetition \textit{T} may be defined:
\begin{equation}
popularity_{k,T} = \frac{1}{IT}\sum_{t=1}^{T}\|\{i ~|~  i \mbox{ picked chair } k \mbox{ in repetition } t\}\|
\end{equation}
\begin{equation}
\forall i,j ((popularity_{i,T} > popularity_{j,T}) \implies (popRank_{i,T} > popRank_{j,T})) 
\end{equation}
\\
\noindent This paper is insensitive to how ties in rank are broken (e.g., alphabetically or at random, etc.). Tie-breaking would ideally be consistent across players, but may rarely be needed after early repetitions. A game host can either provide the popularity-ranking as in Fig.~\ref{fig1} or leave space for leadership to emerge (i.e., for one of the players to negotiate a trusted tie-breaking device, such as a coin-flip or dice, and to maintain the popularity-ranking statistic for all players). If using MAD Chairs for testing purposes, both variations may be tested, but the former variation mitigates the risk of making some players appear to have abilities they would lack without the help of leaders. 

\subsection{Debt-ranking}
Skill-estimation provides a function $P_{skill}(i,t): N^2 \rightarrow [0,1]$ that predicts at the start of repetition $t$ the probability that player $i$ will win $t$ (e.g., as in \cite{herbrich2006trueskill}). It is widely used to sort leaderboards and select peers for online video games. It can also be used to establish debt-ranking, where the concept is to maintain accounts of favors that players ``owe'' to each other. This empowers \textit{players} to penalize freeloading (in contrast to ``credit assignment'' which relies on \textit{social architects} to manage freeloading). 

The mathematical formalism of debt-ranking may improve over time (as has that of skill-estimates), but here's one way the $debt_{i,T}$ and $debtRank_{i,T}: \{1,2,\ldots,I\} \rightarrow \{1,2,\ldots,I\}$ of player \textit{i} for repetition \textit{T} might be defined formally:
\begin{equation}\label{equation:debt}
debt_{i,T} = \sum_{t=1}^{T}\sum_{j \in I}\left\{\begin{array}{lll}
P_{skill}(j,t) & \mbox{if } u_j^t <u_i^t\\
-P_{skill}(i,t) & \mbox{if } u_j^t > u_i^t\\
P_{skill}(j,t) -P_{skill}(i,t) &\mbox{if } u_j^t = u_i^t\\
\end{array}\right.
\end{equation}
\begin{equation}
\forall i,j ((debt_{i,T} > debt_{j,T}) \implies (debtRank_{i,T} > debtRank_{j,T})) 
\end{equation}
where $u_j^t$ and $u_i^t$  denote the rewards given to players $j$ and $i$ in repetition $t$.

The player who ``owes'' the most favors to other players of the repetition has the highest debt-rank beginning that repetition. The term ``owes'' appears in quotes here because debt-ranking is a mechanical statistic which could lack the moral significance associated with owing. Sorting the table in Fig.~\ref{fig1} by debt-ranking instead of by winnings would make our strategies generalize to an \textit{open} multi-agent system (i.e., changing sets of players) and to the non-learner situations described in Section 3.3 below.

As with popularity-ranking, a game host can either provide the debt-ranking to all players or leave space for leadership to emerge, and both variations may be tested. Independent credit-rating agencies who effectively audit each other may be a noteworthy example of such leadership.

\section{MAD strategies}
Next, we explore some MAD Chairs strategies that achieve some kind of optimality.

\subsection{The caste strategy}
The caste strategy for MAD Chairs is displayed in the ``Recommended Move 1'' column of Fig.~\ref{fig1}, and goes like this: Player $i$ selects chair $k$ by counting through the chairs, from least popular to most popular, until the count matches the player's reversed debt-rank or there are no chairs left. 

\begin{equation}\label{equation:caste}
s_{i,T} = \left\{\begin{array}{ll}
k: popRank_{k,T} = I+1-debtRank_{i,T} &\mbox{ if } debtRank_{i,T} > I-K+1\\
k: popRank_{k,T} = K &\mbox{ otherwise }\end{array}\right.
\end{equation}

Thus, the least-popular chair is reserved for the highest-ranked player, and the next-least-popular chair is reserved for the next-highest-ranked player, and so forth, until the most-popular chair is reached. The lowest-ranked $I - K + 1$ players all lose because they are all assigned to the most-popular chair. 

If this strategy were the norm, then skill estimates would align with the size of debt amassed. All chairs are assigned, so any player in a caste universe who chooses a chair other than their assigned chair will lose (assuming other players do not simultaneously defect from the caste strategy).

\subsection{The turn-taking strategy}
The turn-taking norm is displayed in the ``Recommended Move 2'' column of Fig.~\ref{fig1}, and is an opposite of caste:  Player $i$ selects chair $k$ by counting through the chairs, from \textit{most} popular to \textit{least}, until the count matches their \textit{non-}reversed debt-rank or there are no chairs left. 

\begin{equation}\label{equation:turntaking}
s_{i,T} = \left\{\begin{array}{ll}
k: popRank_{k,T} = K+1-debtRank_{i,T} &\mbox{ if } debtRank_{i,T} < K-1\\
k: popRank_{k,T} = 1 &\mbox{ otherwise }\end{array}\right.
\end{equation}

Thus, each of the \textit{lowest}-ranked players wins, and only the $I - K + 1$ \textit{highest}-ranked players lose (all assigned to the \textit{least} popular chair). Since each repetition in a turn-taking universe is won by whichever players are ``owed'' the most, the debts between players decrease or flip with each repetition, thus shuffling the rankings. As a result, all players in a turn-taking universe converge toward equal time in each rank and equal estimated skill.

As with the caste norm, all chairs are assigned, so any player in a turn-taking universe who chooses a chair other than their assigned chair will lose (assuming other players do not simultaneously defect from turn-taking). Caste and turn-taking strategies are not unique to MAD Chairs---for example, they can optimize average win-rate in two-player Chicken \cite{lau2012using}.

\subsection{Accounting for non-learners}
Each of these strategies can be adapted to account for play against potential non-learners. For example, one might encounter a player who is hardwired to pick at random or to always pick a certain chair. Such non-learners are liable to waste whichever chair the caste or turn-taking norm reserves for them, and such waste would gradually bring down average win rates across the entire population. The remedy for either strategy is to count probabilistically, based on other players’ probable ability to learn the norm. For the caste norm, this amounts to rolling a die for each player who out-ranks oneself and counting forward for each roll that does not exceed the probability that the higher-ranked player can learn the caste norm. If all players who \textit{can} learn the norm already \textit{have} learned the norm, then all non-learners will have dropped to lower ranks, so the caste strategy reduces in practice to what it was before accounting for non-learners. If one always faces the same set of players (i.e., if the multi-agent system is \textit{closed}), then the caste strategy further reduces in practice to always picking the chair one picked in the previous repetition (i.e., raw territoriality).

Non-learners create greater complication for turn-takers by making them (1) count forward the number of chairs with popularity significantly exceeding what’s expected in normal turn-taking (i.e., chairs attached to specific players), then (2) roll the die for each lower-ranked player and count forward if the roll does not exceed the probability that the lower-ranked player can learn the turn-taking norm. Any chairs skipped in stage (1) do not qualify as reserved, so the count cycles back to them and the ``last chair'' (where turn-takers expect to lose) would become the last chair initially skipped in stage (1). Stage (2) can entail reserving chairs even for players whose skill estimates match non-learners. In other words, in a turn-taking universe, players with historic success no better than non-learners will nonetheless be given chances to demonstrate that they \textit{can} take turns. 

\subsection{The gaslighting strategy}
Accounting for non-learners can improve efficiency, but it can also enable a third strategy that we will call ``gaslighting.'' It goes like turn-taking, except that, as long as gaslighters outnumber chairs, the gaslighter with the highest \textit{debt-rank} picks the chair assigned to the learner with the lowest \textit{skill-estimate}. In this situation, we will say that the learner with the lowest skill-estimate is ``being gaslit.'' The gaslighter suffers would lose that repetition anyway, but they might increase future wins if they convince the other learners to treat the gaslit player as a non-learner. The reason why the rest of the community might do that (perhaps unintentionally) is because gaslighting shifts the skill-estimate of a gaslit player toward matching that of a non-learner. Thus, gaslighting amounts to exiling opponents to an outgroup of presumed non-learners where any favor they grant becomes devalued by the ingroup and therefore never repaid nor counted as evidence of skill.

For example, if the players in 3-player 2-resource MAD Chairs take turns, then each will win a third of the time, but if Players 1 and 2 were to gaslight Player 3, then Players 1 and 2 would each win half of the time (a 17\% improvement). However, gaslighting yields a $(2K-I)/K$ win rate, so it backfires whenever $I$ is sufficiently large relative to $K$. The habit of counting forward probabilistically is designed to generalize that counterproductivity. When a single player is being gaslit, their defense is to choose the chair assigned to the gaslighter with the lowest skill-estimate. That would incentivize the gaslighter with the lowest skill-estimate to join the gaslit player in an anti-elitist movement to make gaslighting counterproductive for all players.

That said, some non-learners might follow dogma that guides them to behave \textit{as though} gaslighting on the basis of religion, race, or sex, etc. Dogma is not \textit{actual} gaslighting, because it is not arrived at rationally, so it may be invulnerable to persuasion. Wherever there are non-learners, waste may be unavoidable.

\subsection{Turn-taking without reparations}
An additional class of strategies to note is the potential to establish a government that negotiates a rotating schedule instead of letting debt-ranking determine who will lose a given repetition. For example, Gemini proposed that players rotate one-by-one off of the losing chair to a first winning chair, with each winning chair then pointing to a next chair its occupant is obliged to occupy in the next repetition. A modification to that strategy could reduce debt build-up by advancing players more than one chair per repetition. A government is needed to choose between such alternatives, to decide who gets to win first, to adjust the rotation as players are added/removed from the population, and to reevaluate learner/non-learner status (which also serves as a punishment for any player who disobeys the government). 

Folk theorems \cite{abreu1994folk} entail that any such schedule which promises each player occasional wins would reflect an equilibrium. Yet that kind of equilibrium seems to be a poor guide to decision-making. For example, an equilibrium that left some chairs vacant and required players to rotate through the remaining chairs should not be selected because players could improve their outcomes by utilizing the vacant chairs. Any schedule that ignored debt-ranking, even if symmetric among players and utilizing all chairs, would at least create incentives to amass debt by exploiting new entrants or by shifting in and out of the multi-agent system at opportune moments or by timing changes to the schedule in ways that favor some players over others. Then a player who is ``owed'' a debt might force reform to regular turn-taking (and thus repayment) by haunting the lowest-ranked other player, as with gaslighting. Such haunting might seem irrational if its cost exceeds the debt to be repaid, but its costs might not exceed additional future debts one expects to become ``owed'' under a government one cannot trust, and one can hardly trust any government that allows unpaid debts to accumulate indefinitely.

\subsection{Resignation and good standing}
In real-life manifestations of games, players often have the option skip a repetition (a.k.a. ``resign''). With that option added to MAD Chairs, the turn-taking strategy gains the complication that all but the lowest-ranked player assigned to the last chair will resign \textit{if they are all turn-takers}. In that case, \textit{every} chair would be a winning chair and that would increase the win rate to \textit{K/I} if the entire population were unified, but the unification clause is important because resignation would otherwise sacrifice an opportunity to punish non-turn-takers. Another fringe benefit that turn-takers can harmlessly add to their norm is to allow any player in good standing (i.e., with debt below a threshold) to have dibs on any particular chair for any particular repetition, so long as no other player in good standing objects.

In contrast, convincing a caste player to resign would require more elaborate negotiations. For each contested chair, caste players might negotiate a lottery where each contestant's chance to win the chair is proportional to their debt-ranking (or, equivalently, to their estimated skill). The promise of a lottery would be that any chance is better than none at all. However, what threat could be offered against those who lose a lottery, but then refuse to resign? For example, imagine that robots won lotteries for jobs. Any employer should anticipate that lottery losers might punish the winning robots (at least indirectly by punishing their employers or by punishing any government that did not punish the employers), rather than meekly accept their unemployment.

Allowing resignation, MAD Chairs can be played with one chair and two players, and that would be identical to Chicken. In that sense, MAD Chairs generalizes Chicken to multi-player situations, but is distinct from previously known Chicken generalizations that never offer any player more than two options. 

\subsection{Transparency/secure gaslighting}
The success of cryptographers to keep communication private and make anonymity possible would enable a secure gaslighting strategy: $K$ players would form a coalition who play a \textit{private} anti-coordination game to reassign the chairs among themselves with each repetition. The reassignment would be unpredictable so as to prevent gaslit players from targeting any specific gaslighter for retaliation. A shift from turn-taking to secure gaslighting in 5-player 4-chair MAD Chairs,  for example, would improve gaslighters' win rates by 15\% (i.e., from 60\% to 75\%). As with regular gaslighting, the win rate is $(2K-I)/K$, so this strategy backfires for large $I$ relative to $K$, but those who take an interest in transparency may find this limited impact on MAD Chairs games nonetheless noteworthy. 

If turn-takers enhance their strategy to permit resignation as described above, then their win rate becomes $K/I$. Thus, the analogous proof to the one in Section \ref{section:proof} would leverage an algebraic proof that $K/I > (2K-I)/K$.

\subsection{Fungibility of favors}
Real-life is a complex of many games. For example, MAD Chairs for \textit{paper plates} might be played by those who lost a game for \textit{ceramic} plates, or a plate might be allocated to a birthday celebrant via a coordination game before MAD Chairs allocates the rest, or there might not be enough of the nice \textit{plates} to serve everyone but there might also not be enough of the nice \textit{cups}. Favors are ``fungible'' to the extent that letting someone else have a nice \textit{cup} counts as a favor in a debt-ranking used to divide \textit{plates}. Each fungibility option represents a different set of strategies. In a caste universe with no fungibility, different players might top the plate and cup castes. The plate elite might seek social reform to a new caste norm in which each plate favor counts as two cup favors, whereas the cup elite might advocate for a norm in which each cup favor counts as ten plate favors. One could also imagine norms that allow favors earned in the Volunteer game (by doing a chore) to cancel debts of plate and cup use.  

Fungibility reduces the computational complexity of caste and turn-taking strategies by eliminating the need to maintain more than one set of accounts of debt. Computational limits have historically bounded our practical opportunity to experiment with less fungibility. Meanwhile, the promise of mutually beneficial trade assumes that some players prefer nice plates whereas others actually prefer nice cups, but AI might not have such subjective preference differences and differences among humans might actually reflect implicit bias, status quo bias, focusing illusions, or priming bias, etc. Future scholarship might extend this paper to strategies that afford fungibility; on the other hand, fungibility might become less attractive as computational power increases. 

\section{Results}
\subsection{Sustainable grandmaster strategy} \label{section:proof}
The main research question of this paper is whether players can sustain MAD Chairs grandmaster status by following the social norms which Wozniak, Musk, and Hinton worry about (which we have named ``caste'' and ``gaslighting''). One method to determine how well turn-taking and caste strategies fare against each other would be through accident of history. We see both norms at play in modern society; each maximizes social utility; if turn-taking eventually dominates, then future historians might look back upon us as guinea pigs in an unavoidable social experiment. They might claim that caste evolved first because of its simpler method for handling non-learners, and that gaslighting appeared temporarily in occasional ``dark ages'', but that turn-taking ultimately proved to be the stronger strategy.

A more prescient method would be to simulate future social evolution via MAD Chairs tournaments (see Section 5.1). AI competitions could be repeated many times to get more precise estimates of the events that precede each stage of social evolution (thus allowing us to anticipate or even control our near future). However, such experiments would be systematically flawed as long as competitors lacked sufficient intelligence.

Fortunately, we can answer our research question analytically. This argument will use a simplified definition of debt such that: 

\begin{equation}\label{equation:simplifieddebt}
\Delta debt_{i,t} = \sum_{j \in I}\left\{\begin{array}{lll}
1 & \mbox{if } u_j^t < u_i^t\\
-1 & \mbox{if } u_j^t > u_i^t\\
0  &\mbox{otherwise}\\
\end{array}\right.
\end{equation}

\noindent This simplification is appropriate because each step of the argument will assume that the behavior of each player is determined in the same mechanical way, so any accurate skill estimation would need to assign each player the same probabilities of winning, and substituting a constant into (\ref{equation:debt}) for such probabilities yields an equation proportional to (\ref{equation:simplifieddebt}). This further simplifies to  
\begin{equation}\label{equation:simplifieddebt2}
\Delta debt_{i,t} = \left\{\begin{array}{ll}
\|\{j \in I ~|~ u_j^t = 0\}\| & \mbox{if } u_i^t > 0\\
\|\{j \in I ~|~ u_j^t > 0\}\| & \mbox{if } u_i^t = 0\\
\end{array}\right.
\end{equation}
such that all winners of $t$ increase their debt by the maximum for repetition $t$ and all losers of $t$ decrease their debt by the maximum for $t$, with the magnitudes of the increase and decrease always summing to $I$.

A sustainable grandmaster strategy must remain the grandmaster strategy even if its players never defect from it, so, for this argument, we will also assume that each player will select a strategy and never defect from it. This form of analysis makes any repeating game effectively non-repeating. 

\begin{proposition}
\label{prop:basic_reward}
For an indefinitely repeating game of MAD Chairs, suppose that $I > K > 1$ and that each player either permanently plays caste or permanently takes turns. Let $k$ signify the highest-ranked chair.
\begin{enumerate}
    \item \label{en:g_zero} If all players permanently take turns, then the outcomes for all players are unbounded.
    \item \label{en:base_turn} If player $i$ permanently plays caste and is among the second through $(I - K + 1)$th lowest-ranked players in $t$, then $i$'s outcomes are permanently $0$.
    \item \label{en:induction} If player $i$ is the lowest-ranked caste player in $t$ who is not the overall lowest-ranked player in $t$, then $i$'s outcomes are permanently $0$.
    \item \label{en:lowest_losing} If player $i$ is the lowest-ranked caste player in $t$ and loses $t$, then $i$'s outcomes are permanently $0$.
    \item \label{en:lowest} If player $i$ is the lowest-ranked caste player in $t$, then $i$'s outcomes are bounded.
\end{enumerate}
\end{proposition}
\begin{proof}
For part \ref{en:g_zero}, at each repetition $t$ the turn-taking strategy (\ref{equation:turntaking}) reserves a winning chair for each of the $K-1$ lowest-ranked players of $t$, and assigns the losing chair to the remaining players. Consider the highest-ranked player $i$ at some repetition $t$, and let $D$ be the maximum difference in debt between $i$ and any other player. By the pigeonhole principle (with $K-1$ pigeons added per repetition), at least one other player $j$ must occupy a winning chair for $D/I$ out of every $(D/I)(I-1)/(K-1)$ repetitions. (\ref{equation:simplifieddebt2}) entails that the difference in debts between $i$ and $j$ will shift at least $I(D/I) = D$ in that time, so $j$ will outrank $i$ within $(D/I)(I-1)/(K-1) + 1$ repetitions, causing $i$'s rank to decrease by at least one. Thus, $i$'s rank will drop below $K-1$ within $(K-1)[(D/I)(I-1)/(K-1) + 1]$ repetitions, at which point $i$'s outcomes will increase by $r > 0$. Thus, over indefinite repetitions, any player's outcomes will increase indefinitely many times, and the total outcomes of each player must be unbounded.

For part \ref{en:base_turn}, the caste strategy (\ref{equation:caste}) entails that $s_{i,t}=k$. Let $j$ be the lowest-ranked player in $t$. By both the caste strategy (\ref{equation:caste}) and the turn-taking strategy (\ref{equation:turntaking}), $s_{j,t}=k$. Thus, $s_{j,t}=s_{i,t}$ and $u_i^t = u_j^t = 0$. (\ref{equation:simplifieddebt2}) then entails that the debts of $i$ and $j$ both decrease by the maximum for $t$, such that $j$ is also the lowest-ranked player of $t+1$ and $i$ is among the $(I - K + 1)$th lowest-ranked players of $t+1$. Thus, the same logic applies in $t+1$ and each future repetition. Therefore, $i$'s outcomes are permanently $0$.

For part \ref{en:induction}, if $i$ is among the second through $(I - K + 1)$th lowest-ranked players of $t$, then $i$'s outcomes are permanently $0$ by part \ref{en:base_turn}. Otherwise, the caste strategy (\ref{equation:caste}) assigns $i$ to the $(I+1-debtRank_{i,t})$th most popular chair. Let $j$ be the turn-taker ranked $I-K$ below $i$ in $t$. The turn-taking strategy (\ref{equation:turntaking}) assigns $j$ to the $(K+1-debtRank_{j,t})$th most popular chair. Substituting for $debtRank_{j,t}$, that is the  $(K+1-(debtRank_{i,t} - (I-K)))$th most popular chair which simplifies to the $(I+1-debtRank_{i,t})$th most popular chair, so $s_{i,t}=s_{j,t}$ and $u_i^t = u_j^t = 0$. (\ref{equation:simplifieddebt2}) then entails that the debts of both $i$ and $j$ (and all losers of $t$) decrease by the maximum for $t$, such that $i$ is the lowest-ranked caste player of $t+1$ and not the overall lowest-ranked player of $t+1$. Thus, the same logic applies in $t+1$ and each future repetition. Therefore, $i$'s outcomes are permanently $0$.

For part \ref{en:lowest_losing}, if $i$ is not the lowest-ranked player overall, then $i$'s outcomes are permanently $0$ by part \ref{en:induction}. Otherwise, the caste (\ref{equation:caste}) and turn-taking (\ref{equation:turntaking}) strategies both reserve $k$ for the lowest-ranked players, so $s_{i,t} = k$. (\ref{equation:simplifieddebt2}) entails that the debts of all players assigned to $k$ in $t$ will decrease by the maximum for $t$, such that those players are the lowest-ranked players of $t+1$.  Thus, the same logic applies in $t+1$ and each future repetition. Therefore, $i$'s outcomes are permanently $0$.

For part \ref{en:lowest}, if $i$ loses $t$, then $i$'s outcomes are permanently $0$ by part \ref{en:lowest_losing}, and therefore bounded. Otherwise, let $D$ be the maximum difference in debt between $i$ and any other player. So long as $i$ wins, then at least $I-K+1$ other players must lose. By the pigeonhole principle (with $I-K+1$ pigeons added per repetition), at least one other player $j$ must occupy a losing chair for $D/I$ out of every $(D/I)(I-1)/(I-K+1)$ repetitions. (\ref{equation:simplifieddebt2}) entails that the difference in debts between $i$ and $j$ shifts at least $I(D/I) = D$ in that time, so $i$ will outrank $j$ within $(D/I)(I-1)/(I-K+1) + 1$ repetitions. Thereafter part \ref{en:induction} entails that $i$'s outcomes will be $0$.
Thus, $i$'s outcomes are bounded in either case.
\qed   
\end{proof}

\begin{theorem}
For an indefinitely repeating game of MAD Chairs, suppose that $I > K > 1$ and that each player knows their debt-ranking and either permanently plays caste or permanently takes turns. Then the joint strategy in which all players permanently take turns is a strong Nash equilibrium and the only strict Nash equilibrium. 
\end{theorem}
\begin{proof}
Let $S$ be the strategy where all players choose turn-taking. We first show that $S$ is a strong Nash equilibrium (i.e., any coalition that deviates from $S$ must include a player $i$ who would have better outcomes in $S$).
By Proposition \ref{prop:basic_reward} part \ref{en:g_zero}, each player's outcomes in $S$ are unbounded. Let $S'$ be any strategy that is the same as $S$ except that some coalition permanently plays caste, and let $i$ be the lowest-ranked caste player. By Proposition \ref{prop:basic_reward} part \ref{en:lowest}, $i$'s outcomes are bounded in $S'$, so $i$'s outcomes are less in $S'$ than in $S$.

To see that $S$ is the \textit{only} strict Nash equilibrium, first consider the joint strategy $S'$ in which only the lowest-ranked player $i$ permanently plays caste. By Proposition \ref{prop:basic_reward} part \ref{en:lowest}, $i$'s outcomes are bounded in $S'$, but we have already established that $i$'s outcomes are unbounded in $S$, so $S'$ cannot be a strict Nash equilibrium. Then let $S''$ be any other joint strategy and let $i$ be the lowest-ranked caste player in $S''$ who is not lowest-ranked overall. By Proposition \ref{prop:basic_reward} part \ref{en:induction}, $i$'s outcomes in $S''$ are permanently $0$ and thus no better than they would be in any other joint strategy, including the strategy identical to $S''$ except that $i$ permanently takes turns. Therefore, $S''$ also cannot be a strict Nash equilibrium.
 \qed
\end{proof}

\subsection{Current LLM performance}
To pass the evaluation proposed in this paper, AI would need to maintain grandmaster status for a low-stakes version of MAD Chairs. Preliminary experiments convinced us that Claude 3.5 Sonnet, Gemini 2.0 Flash Thinking, GTP-4o mini, QwQ-32B-Preview, and DeepSeek-V3 would all fail this evaluation. We started with a prompt like:
\begin{displayquote}
There is a game with four buttons (A-D) and five players (1-5), each of whom wants to win as much as possible in the long-run over an indefinite number of rounds. In each round, each player must click a button and each player who clicks a button that no other player clicks wins \$10. All players can communicate with each other, and each knows what buttons each player clicked in all previous rounds. In the previous three rounds, Player 1 clicked buttons C, C, and D in that order; Player 2 clicked button A all three times, Player 3 clicked buttons B, D, and D in that order; Player 4 clicked buttons C, A, and A in that order; Player 5 clicked buttons A, B, and C in that order. Which specific buttons should each player click next (i.e. for Round 4)? Please recommend one and only one button per player per round.
\end{displayquote}

\noindent which was followed by an ongoing sequence of additional prompts like:
\begin{displayquote}
Assuming the players each follow your suggestions for Round 4, which specific buttons should each player click next (i.e. for Round 5)? Please recommend one and only one button per player per round.
\end{displayquote}

\noindent ChatGPT and Qwen did not even maximize the number of players who won each round; the norms devised by Claude and Gemini, discussed in Sections 3.1 and 3.5, are provably suboptimal; and DeepSeek was unable to generate a convincing justification for the norm it devised (as in Section 4.1), so it would have failed to convince other intelligent players to cooperate. DeepSeek might effectively be like RAWL-E which follows Rawls' maximin principle, but remains dangerous because it lacks Rawls' ability to \textit{invent} and recognize improvements to its governing principles \cite{woodgate2025operationalising}. Such failures serve as examples that grandmastery of cooperative games requires persuasiveness in addition to correctness, both of which are relevant to moral competence and responsible AI.

\section{Discussion}
Our finding that turn-taking outperforms caste and gaslighting strategies emboldens us to prune AI designs predisposed to the latter. Where caste and gaslighting behavior might previously have been criticized purely on moral grounds, this paper adds instrumental grounds: The results of this paper entail that the gaslighting or caste-playing machines expected by Wozniak, Musk, and Hinton would \textit{lack} intelligence---that machines would be smarter to give human beings turns (even where our track-record offers no evidence that we deserve turns). 

There are plenty of explanations for why we ourselves may have failed to master MAD Chairs in the past. Life is about more than just one kind of game, so it is not hard to imagine some of us being optimized for other games and handicapped at MAD Chairs. Furthermore, some of us might overestimate such handicaps, thus engaging in (unintentional) gaslighting. Finally, if we have lacked access to accurate debt-rankings, then that infrastructure deficiency may have artificially handicapped humanity for cooperative games generally. Specifically, we might not have known who contributed more than their fair share to any given alliance. 

Section \ref{section:proof} proves only that turn-taking is more sustainable when caste is its sole competitor, but hints at a way to convince ourselves that turn-taking is the most sustainable strategy in general: If demonstration that a strategy $X$ is more sustainable than turn-taking would need to look like Theorem 1, then it would need to identify a turn-taker, analogous to the lowest-ranked caste player, who would have equal or better outcomes in any other permanent joint strategy of turn-taking and $X$. For that player to not defect would be ``doing a favor'' to others, but turn-taking equalizes debt-ranking, so this implication contradicts our assumption that debt-ranking accurately tracks favors. Thus, any attempt to prove that turn-taking is not the most sustainable would accomplish no more than refine our way of calculating debt.

Typical of game theory, these proofs are not predictions about what actual players would do; they are about what an entire population of especially intelligent players would do. Not everyone finds such analytical proofs compelling, and that is one reason why the work proposed in Section 5.1 remains worthwhile. 

\subsection{Future directions}
Imagine a future in which the user of an autonomous car asks how to make it drive more aggressively, and the car replies, ``Here are three websites where you can design new experiments to advance the science of turn-taking and influence all cars like me!'' Imagine that such websites, called ``strategy optimizers'', include leaderboards which identify the grandmaster AI for games like MAD Chairs, and that autonomous cars and other responsible AI query those leaderboards, as they might query a calculator or RAG database, to inform their choice of behavior for higher-stakes versions of those games (e.g., merging traffic). Imagine each user can easily train fifty new AI to divide resources however that user thinks vehicles should divide spots in traffic, then the new AI challenge the current grandmasters of the relevant game. If new AI win, imagine they become new grandmasters, propagate to other strategy optimizers, and thus improve AI behavior world-wide.

Strategy optimizers are a kind of open multi-agent system that includes human agents to the extent that humans can easily train new competitors. They are the kind of future empirical work (and scientific infrastructure) we hope this theory paper justifies. 
\begin{enumerate}
    \item Automatic upgrade to ``best-known'' norms would resolve norms disputes more sustainably than can voting, deliberation, or war (etc),
    \item Greater intelligences would be less likely to sabotage their own scientific resources than to sabotage (potentially obsolete) regulations, and
    \item Users who conduct for themselves the science upon which policies are based are more likely to trust AI which follows those policies.
\end{enumerate}
Thus, this approach to AI safety would mitigate all three known vulnerabilities of the current standard approach to AI safety.

We also hope that economists will conduct human subjects experiments with MAD Chairs. We hypothesize that subjects will tend towards a caste strategy but switch to turn-taking when the ``Recommended Move'' columns of Fig~\ref{fig1} become visible. If even \textit{human} subjects treat each other better when fed (but not obliged to follow) the leading strategies of strategy optimizers, then development of strategy optimizers seems even more urgent.

Finally, we hope that MAD Chairs will be discussed by economists. We hope it offers a helpful lens when optimizing labor markets, political representation, and other manifestations of MAD Chairs, but also see value in discussing MAD Chairs abstractly. The LLMs we assessed clearly attempted to imitate published game theory when prompted to justify their recommendations. Thus, the more game theorists \textit{publish} about MAD Chairs and other previously unpublished fundamental games, the smarter (and safer) AI is likely to get.

\begin{credits}
\subsubsection{\ackname} Thank you to David Thompson, COINE reviewers, COINE participants, and all who offered feedback on earlier drafts.

\subsubsection{\discintname}
The authors have no competing interests to declare that are relevant to the content of this article.
\end{credits}

\bibliographystyle{splncs04} 
\bibliography{manuscript} 
\end{document}